\documentclass[runningheads]{llncs}
\usepackage{graphicx}
\usepackage{caption}
\usepackage{subcaption}
\usepackage{amsmath}
\usepackage{array}
\usepackage{amsfonts}
\usepackage{booktabs}
\usepackage{siunitx}
\usepackage{float}
\usepackage{ragged2e}
\usepackage[colorlinks=true]{hyperref}

\DeclareMathOperator{\Fr}{\mathcal{F}}
\DeclareMathOperator{\La}{\mathcal{L}}
\DeclareMathOperator{\Me}{\mathcal{M}}
\DeclareMathOperator{\D}{\mathcal{D}}
\newcolumntype{P}[1]{>{\centering\arraybackslash}p{#1}}
\newcolumntype{M}[1]{>{\centering\arraybackslash}m{#1}}
\begin{document}
\title{Knowledge-Guided Multiview Deep Curriculum Learning for Elbow Fracture Classification}
\titlerunning{Knowledge-guided Multiview Deep Curriculum Learning}
\author{Jun Luo\inst{1} \and
Gene Kitamura\inst{2} \and
Dooman Arefan\inst{2} \and
Emine Doganay\inst{2} \and
Ashok Panigrahy\inst{2,3} \and
Shandong Wu\inst{1,2,4}}
\authorrunning{J. Luo et al.}
\institute{Intelligent Systems Program, School of Computing and Information, University of Pittsburgh, Pittsburgh, PA, USA \\
\email{jul117@pitt.edu} \and
Department of Radiology, School of Medicine, University of Pittsburgh, Pittsburgh, PA, USA \and
University of Pittsburgh Medical Center Children's Hospital of Pittsburgh, Pittsburgh, PA, US \and
Department of Biomedical Informatics and Department of Bioengineering, University of Pittsburgh, PA, USA \\\email{wus3@upmc.edu}}
\maketitle              % typeset the header of the contribution
\begin{abstract}
Elbow fracture diagnosis often requires patients to take both frontal and lateral views of elbow X-ray radiographs. In this paper, we propose a multiview deep learning method for an elbow fracture subtype classification task. Our strategy leverages transfer learning by first training two single-view models, one for frontal view and the other for lateral view, and then transferring the weights to the corresponding layers in the proposed multiview network architecture. Meanwhile, quantitative medical knowledge was integrated into the training process through a curriculum learning framework, which enables the model to first learn from “easier” samples and then transition to “harder” samples to reach better performance. In addition, our multiview network can work both in a dual-view setting and with a single view as input. We evaluate our method through extensive experiments on a classification task of elbow fracture with a dataset of 1,964 images. Results show that our method outperforms two related methods on bone fracture study in multiple settings, and our technique is able to boost the performance of the compared methods. The code is available at \url{https://github.com/ljaiverson/multiview-curriculum}.

\end{abstract}
\section{Introduction}
Human’s cognitive ability relies deeply on integrating information from different views of the objects. This is particularly the case for elbow fracture diagnosis where patients are often required to take both the frontal view (i.e. Anterior-Posterior view) and lateral view of elbow X-ray radiographs for diagnosis. This is because some fracture subtypes might be more visible from a certain perspective: the frontal view projects the distal humerus, the proximal ulna and the radius~\cite{whitley2015clark,el1997acute,stevens1999imaging}, while the lateral view shows the coronoid process and the olecranon process~\cite{whitley2015clark,goldfarb2012elbow,sandman2016effect}. In practice, it is also common that some patients only have a single view radiograph acquired, or have a missing view for various reasons.

In recent years, the advance of deep learning has been facilitating the automation of bone fracture diagnosis~\cite{kalmet2020deep,cheng2021scalable,guan2020arm} through multiple views of X-ray images, which shows faster speed and decent accuracy compared to human experts~\cite{kitamura2019ankle,rayan2019binomial,krogue2020automatic}. However, few methods leverage multiview information, which provide more visual information from different perspectives for elbow fracture diagnosis.

In this work, we propose a novel multiview deep learning network architecture for elbow fracture subtype classification that takes frontal view and lateral view elbow radiographs as input. While the proposed model is a dual-view (frontal and lateral) architecture, it is flexible as it does not strictly require a dual-view input during inference. Furthermore, our training strategy for the multiview model takes advantage of transfer learning by first training two single-view models, one for frontal view and the other for lateral view, and then transferring the trained weights to the corresponding layers in the proposed multiview network architecture. In addition, we investigate the utilities of integrating medical knowledge of different views into the training via a curriculum learning scheme, which enables the model to first learn from “easier” samples and then transition to “harder” samples to reach better performance.

\begin{figure}[H]
     \centering
     \begin{subfigure}[b]{0.15\textwidth}
         \centering
         \includegraphics[width=\textwidth]{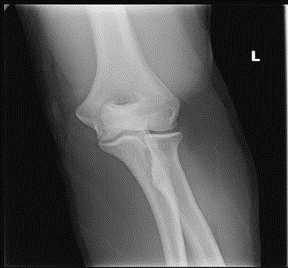}
         \caption{}
         \label{normalf}
     \end{subfigure}
     \hfill
     \begin{subfigure}[b]{0.15\textwidth}
         \centering
         \includegraphics[width=\textwidth]{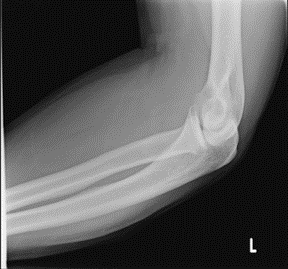}
         \caption{}
         \label{normall}
     \end{subfigure}
     \hfill
     \begin{subfigure}[b]{0.15\textwidth}
         \centering
         \includegraphics[width=\textwidth]{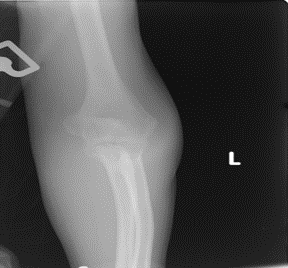}
         \caption{}
         \label{ulnarf}
     \end{subfigure}
     \hfill
     \begin{subfigure}[b]{0.15\textwidth}
         \centering
         \includegraphics[width=\textwidth]{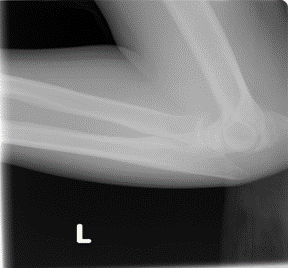}
         \caption{}
         \label{ulnarl}
     \end{subfigure}
     \hfill
     \begin{subfigure}[b]{0.15\textwidth}
         \centering
         \includegraphics[width=\textwidth]{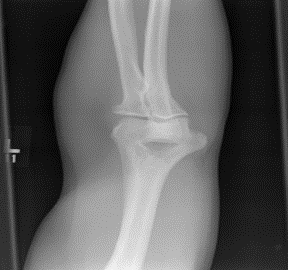}
         \caption{}
         \label{radialf}
     \end{subfigure}
     \hfill
     \begin{subfigure}[b]{0.15\textwidth}
         \centering
         \includegraphics[width=\textwidth]{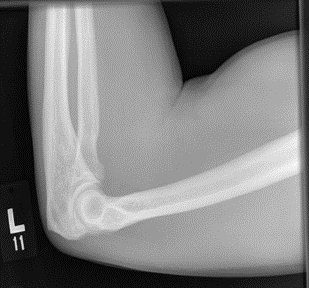}
         \caption{}
         \label{radiall}
     \end{subfigure}
        \caption{Example images from the three categories from our dataset for classification task: (a) and (b) show the frontal and lateral non-fracture category respectively; (c) and (d) show the frontal and lateral ulnar fracture category respectively; (e) and (f) show the frontal and lateral radial fracture category respectively.}
        \label{elbowfracturesubtypes}
\end{figure}

To evaluate our method, we conduct experiments on a classification task of three classes of elbow fractures that shown in Figure \ref{elbowfracturesubtypes}. We compare our method to multiple settings including the single-view models, different combinations of the transfer learning strategy and the knowledge-guided curriculum learning. Our method is also compared to a previous method~\cite{jimenez2019medical}. Results show that our proposed method outperforms the compared methods, and our method functions seamlessly on a multiview and a single-view settings.

\section{Related Work}
Multiview learning~\cite{xu2013survey} takes advantage of data with multiple views of the same objects. Co-training~\cite{blum1998combining,nigam2000analyzing,sindhwani2005co} style algorithms were a group of traditional multiview learning algorithms originally focusing on semi-supervised learning, where multiple views of data were iteratively added to the labeled set and learned by the classifier. Another group of multiview learning algorithms explore Multiple Kernel Learning (MKL), which was originally proposed to restrict the search space of kernels~\cite{NIPS2009_e7f8a7fb,duffy2000leveraging}. Recent work on multiview learning based modeling shows promising effects for medical fields such as bone fracture and breast cancer detection~\cite{kitamura2019ankle,rayan2019binomial,geras2017high}.

Curriculum learning is also an area of active research. It was first introduced by Bengio et al. in~\cite{bengio2009curriculum} to enable the machine learning to mimic human learning by training a machine learning model first with "easier" samples and then transition to "harder" samples. Some existing work focus on integrating domain knowledge into the training process through curriculum learning. For example,~\cite{jimenez2019medical,luo2021medical} integrate domain knowledge by using the classification difficulty level of different classes.

\section{Methods}
\subsection{Multiview Model Architecture}
\begin{figure}
    \centering
    \includegraphics[width=0.7\textwidth]{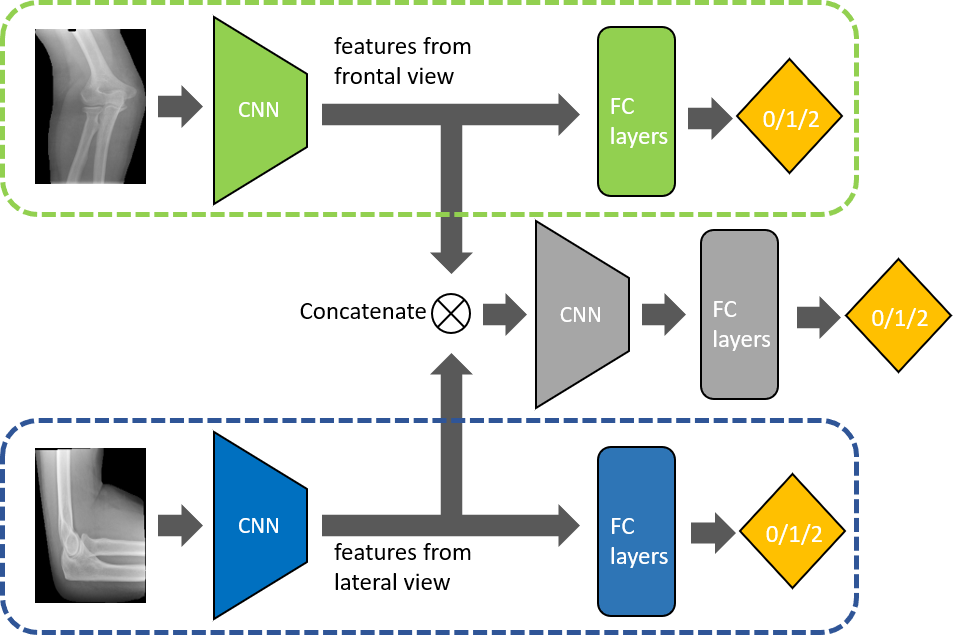}
    \caption{The proposed multiview model architecture. The green and blue dotted line box represent the frontal and lateral view modules, respectively. Yellow diamonds are the predicted labels, 0, 1, 2 corresponding to non-fracture, ulnar fracture, radial fracture respectively} \label{multiview-architecture}
\end{figure}
To incorporate information from both frontal and lateral view for the elbow X-ray images while maintaining the flexibility of being able to output predictions with one view as input, we propose a novel multiview model architecture shown in Figure \ref{multiview-architecture}. In this architecture, during training, pairs of frontal and lateral view images are fed into their corresponding modules for feature extraction by the convolutional neural networks (CNNs). After the feature extraction, the model splits into three branches as shown in Figure \ref{multiview-architecture}. The top and bottom branches take the corresponding single-view features to the fully connected (FC) layers for classification, while the middle branch takes the concatenated features from both views as input to further extract features and then conducts classification.

Consider a data sample triplet $\D_i=\{ x_i^{( F )} , x_i^{( L )}, y_i \}$ where $\D_i$ represents the $i$-th data sample, $x_i^{( F )}$, and $x_i^{( L )}$ are its images from the frontal and lateral view, and $y_i \in \{ 0, 1, 2\}$ is its ground truth label with 0, 1, 2 corresponding to non-fracture, ulnar fracture, radial fracture respectively. We denote the three predicted labels from the three branches of our multiview model as $\Fr( x_i^{( F )} )$, $\La( x_i^{( L )} )$, and $\Me( x_i^{( F )}, x_i^{( L )} )$, where $\Fr$, $\La$, $\Me$ represent the \textit{frontal view module}, the \textit{lateral view module}, and the ``\textit{merge module}'' that contains the two CNN blocks from the frontal and lateral module, the CNN as well as the FC layers in the middle branch. 

During training, we minimize the objective function over the $i$-th data sample computed by equation (\ref{loss}) where $\theta$, $\theta_{\Fr}$, $\theta_{\La}$, and $\theta_{\Me}$ represent the parameters in the entire model, the frontal view module, the lateral view module, and the merge module. As shown in equation (\ref{loss}) (with $C$ being the number of classes), for each module, the loss is computed with cross entropy loss over the corresponding predicted label and ground truth $y_i$ in a one-hot representation.
\begin{equation}
    \begin{split}
        J_{\theta}(  x_i^{( F )} , x_i^{( L )}, y_i ) = J_{\theta_{\Fr}}( x_i^{( F )}, y_i ) + J_{\theta_{\La}}( x_i^{( L )}, y_i ) + J_{\theta_{\Me}}( x_i^{( F )}, x_i^{( L )}, y_i )\\
        = -\sum_{c=1}^C \left( y_{i,c} \left( \log(\Fr( x_i^{( F )} )_c) + \log(\La( x_i^{( L )} )_c) + \log(\Me( x_i^{( F )}, x_i^{( L )} )_c) \right)\right)
    \end{split}
    \label{loss}
\end{equation}

During test phase, if a frontal view image and a lateral view image are both presented, the default final predicted label is the one predicted from the merge module, i.e. $\Me( x_i^{( F )}, x_i^{( L )} )$. Alternatively, if there is only one view, the model will still output a predicted label from the module of the corresponding view credited to the designed architecture of our model.

\subsection{Transfer learning from pretrained single-view models}

\begin{figure}[t]
    \centering
    \includegraphics[width=0.8\textwidth]{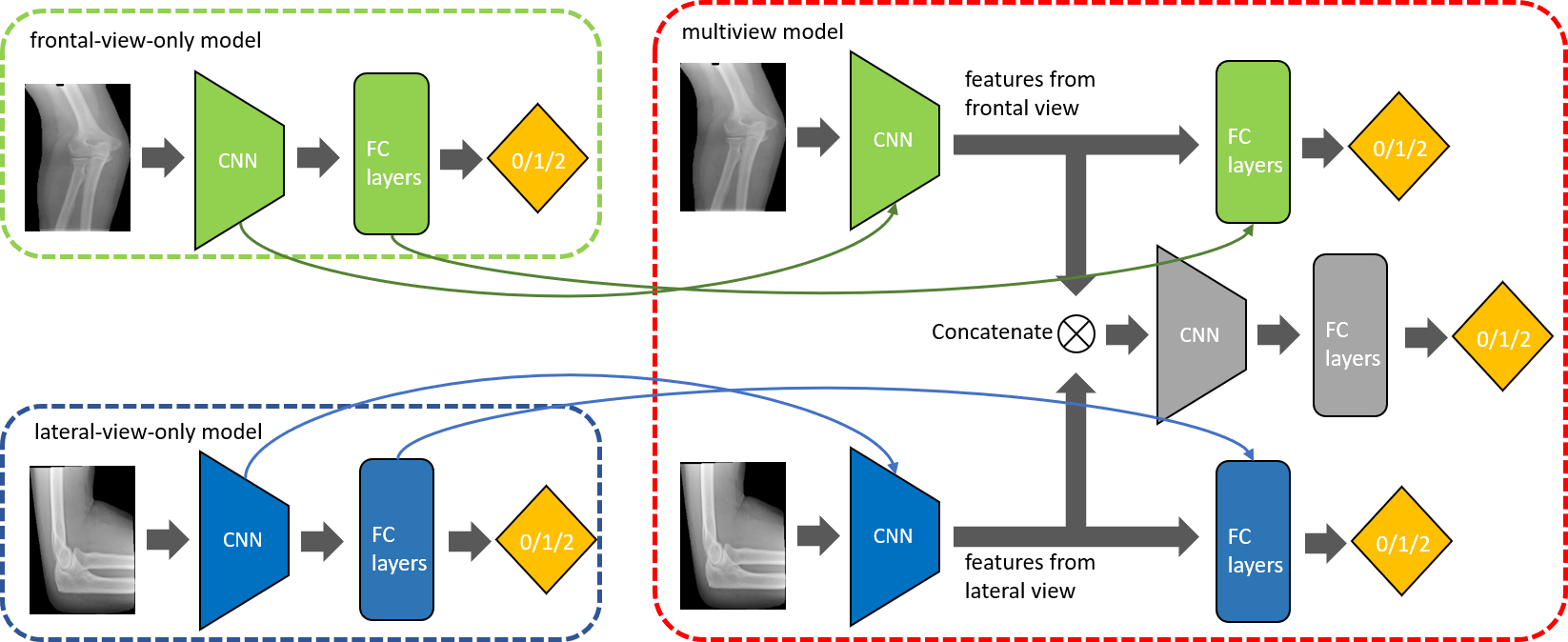}
    \caption{Transfer learning from pretrained single-view models.} \label{trans}
\end{figure}

In most medical applications with deep learning, researchers use the ImageNet~\cite{deng2009imagenet} pretrained model as a way of transfer learning. However, a great number of deep learning models do not have publicly available pretrained weights, especially for self-designed models. Here, we investigate a homogeneous way of transfer learning as shown in Figure \ref{trans}: we first train two single-view models (using the same training set as the one for the multiview model) that have identical structure as the frontal view and lateral view module in the multiview architecture. Then, we transfer the trained weights of the CNNs and FC layers from the single view models to the counterparts of the multiview model (refer to the links in Figure \ref{trans}). For the middle branch (the gray CNN and LC layers blocks in Figure \ref{multiview-architecture}) in the merge module, we randomly initialize their weights. We make all weights trainable in the multiview model.

\subsection{Knowledge-guided curriculum learning}
For the model training, we propose a knowledge-guided curriculum learning to enhance learning effects. The idea of curriculum learning is to enable the training process to follow an ``easy-to-hard'' order, where the easier samples will be fed into the model for training earlier than the harder samples. To do so, we implemented a multiview-based curriculum learning by adapting the method from~\cite{luo2021medical}. We quantify and integrate medical knowledge by scoring the classification difficulty levels of each category of elbow fracture with board-certified radiologist's expertise. Table \ref{scores} shows the quantitative scores reflecting the classification difficulty based on experience of expert radiologists. Note that we use the ``Both views'' scores to train the multiview model, and use ``Frontal/Lateral view only'' for homogeneous transfer learning.
\begin{table}[H]
    \caption{Quantitative classification difficulty levels for each category of elbow fracture (1-hardest; 100-easiest), which enables the integration of medical knowledge into curriculum learning.}\label{scores}
    \centering
    \begin{tabular}{lM{4cm}M{2.5cm}M{2.5cm}}
    \toprule
     &  Non-fracture (normal) & Ulnar fracture & Radial fracture\\ \midrule
    Frontal view only &  30 & 30 & 30\\
    Lateral view only &  35 & 60 & 45\\
    Both views & 45 & 65 & 55 \\ \bottomrule
    \end{tabular}
\end{table}
These scores are used to initialize the sampling probability for each training data point according to equation (\ref{update}) with $e=1$, where $p_i^{(1)}$ is the initial sampling probability for data point $\D_i$, $s_i$ is its score, $s_k$ is the score of the data point $\D_k$, and $N$ is the number of data points in the dataset. Using the sampling probabilities, at the beginning of every epoch, we permute the training set by sampling all the data points without replacement.

\begin{equation}
    \label{update}
    p_i^{(e)}=
    \begin{cases}
        \frac{s_i}{\sum_{k=1}^{N} s_k} & e = 1,\\
        p_i^{(e-1)} \cdot \sqrt[E']{\frac{1/N}{p_i^{(0)}}} & 2 \leq e \leq E',\\
        1/N & E' < e \leq E
    \end{cases}
\end{equation}
This enables the easier samples to have a higher chance of being presented before the harder samples. This chance will be exponentially reduced by updating the sampling probabilities for each data point according to equation (\ref{update}). In this equation, $e$ is the current epoch, $E'$ is the last epoch that we update the sampling probabilities. For the rest of the training ($E' < e \leq E$) the sampling probabilities will be fixed to $1/N$.

\section{Experiments and Results}
\subsection{Experiment settings}
\subsubsection{Dataset and Implementation Details.} This study includes a private dataset of 982 subjects of elbow fractures in an Institutional Review Board-approved retrospective study. The subjects are categorized into three classes: 500 non-fracture (normal) cases, 98 ulnar fracture cases, and 384 radial fracture cases. Each subject includes one frontal and one lateral elbow X-ray image, which makes it a total of 1,964 elbow X-ray images. To increase the robustness of our results, we conduct 8-fold cross validation. For each split of the entire dataset, one fold was used as the hold-out test set. Within the remaining seven folds, we randomly select one fold as the validation set for hyperparameter tuning. The remaining folds are used as the training set. All separations of the dataset are in a stratified manner, which maintains the ratio over different classes. The reported results are averages over the 8 disjoint held-out test sets.

\textit{VGG16}~\cite{simonyan2014very} is used as the backbone for the two single-view models, and the frontal and lateral modules in the multiview model. We customize the middle branch two $3\times3\times512$ convolutional layers with max pooling layers, followed by VGG16's classifier for the FC layers. The hyperparameters are selected based on the best validation AUCs. We use the following hyperparameters for the proposed model: batch size 64, learning rate $10^{-4}$ for the Adam optimizer, and after 16 epochs every sample is treated as having an equal difficulty score. All models were trained on an NVIDIA Tesla V100 GPU. The code is available at \url{https://github.com/ljaiverson/multiview-curriculum}.

\subsubsection{Metrics.} The metrics for the 3-class classification task include accuracy and area under receiver operating characteristic curve (AUC). We also compute a balanced accuracy by averaging the ratios between the number of true positives and the total number of samples with respect to each class, which reduces the effect induced by data imbalance. In addition, we evaluate the models' overall ability to distinguish fracture against non-fracture images. This is done by binarizing the ground truth and predicted labels by assigning 0 to them if they originally are 0, and assigning 1 otherwise. We compute the binary task accuracy and the AUC as two additional measures.

\begin{table}[t]
    \caption{Model performance with both views. The bold numbers correspond to the highest value for each metric (TL: proposed transfer learning from single view models; CL: proposed knowledge-guided curriculum learning).}\label{bothviewsresults}
    \centering
    \begin{tabular}{wl{4cm}M{1.5cm}M{1.5cm}M{1.5cm}M{1.5cm}M{1.5cm}}
    \toprule
    Model  &  Accuracy & AUC & Balanced accuracy & Binary task accuracy & Binary task AUC\\ \midrule
    Single-view-frontal & 0.683 & 0.807 & 0.570 & 0.732 & 0.813 \\
    Single-view-lateral & 0.856 & 0.954 & 0.807 & 0.895 & 0.959 \\ \midrule
    Multiview & 0.854 & 0.958 & 0.796 & 0.884 & 0.964 \\
    Multiview + TL & \textbf{0.891} & 0.966 & 0.847 & \textbf{0.916} & 0.973 \\
    Multiview +~\cite{jimenez2019medical} & 0.818 & 0.939 & 0.746 & 0.864 & 0.952 \\
    Multiview +~\cite{jimenez2019medical} + TL & 0.870 & 0.961 & 0.811 & 0.898 & 0.973 \\
    Multiview + CL & 0.889 & 0.970 & 0.847 & 0.908 & \textbf{0.978} \\
    Multiview + CL + TL &  0.889 & \textbf{0.974} & \textbf{0.864} & 0.910 & 0.976 \\ \bottomrule
    \end{tabular}
\end{table}

\subsection{Results}
As shown in Table \ref{bothviewsresults}, we compare our proposed multiview model with curriculum learning method (CL) and transfer learning (TL) with the following six types of models: 1) two single-view models (frontal/lateral view only), referred as Single-view-frontal/lateral; 2) multiview model with regular training, referred as Multiview; 3) multiview model with only transfer learning strategy, referred as Multiview + TL; 4) multiview model with a previous curriculum training method~\cite{jimenez2019medical}, referred as Multiview +~\cite{jimenez2019medical}; 5) multiview model with \cite{jimenez2019medical} and our proposed transfer learning strategy, referred as Multiview +\cite{jimenez2019medical} + TL; and 6) multiview model with only our curriculum learning method, referred as Multiview + CL. We use the output from the middle branch, as the predicted label.

Attributed to the multiple branches of our model and the customized loss function, our model has the flexibility of generating the prediction with a single view as input. In Table \ref{missingview}, we show the results of the performance from the frontal view module and lateral view module separately. Different from~\cite{jimenez2019medical}, our curriculum updates the difficulty score of every sample after every epoch, which benefits the multiview model. Table \ref{bothviewsresults} shows that with both views presented in the test phase, our method achieves the highest AUC and balanced accuracy with a margin of up to 0.118 compared to the state-of-the-art performance. In settings with missing views, however, our strategy does not always perform the best. Table \ref{missingview} shows that with frontal view as the only input view, our method outperforms all the compared methods per each metric, but with the lateral view as the only input view, our method achieves slightly lower performance than the best results.

\begin{table}[t]
    \caption{Model performance with a single view as input}\label{missingview}
    \centering
    \begin{tabular}{wl{3.5cm}M{1.15cm}M{1.4cm}M{0.8cm}M{1.5cm}M{1.4cm}M{1.5cm}}
    \toprule
    Model & Input view & Accuracy & AUC & Balanced accuracy & Binary task accuracy & Binary task AUC\\ \midrule
    Single-view  & frontal & 0.720 & 0.828 & 0.593 & 0.761 & 0.844 \\
    Single-view + CL~\cite{luo2021medical}  & frontal & 0.683 & 0.807 & 0.570 & 0.732 & 0.813 \\ \midrule
    Multiview & frontal & 0.658 & 0.749 & 0.514 & 0.702 & 0.766 \\
    Multiview + TL & frontal & 0.738 & 0.827 & 0.617 & 0.774 & 0.829 \\
    Multiview +~\cite{jimenez2019medical} & frontal & 0.566 & 0.675 & 0.396 & 0.575 & 0.648 \\
    Multiview +~\cite{jimenez2019medical} + TL & frontal & 0.737 & 0.815 & 0.605 & 0.773 & 0.831 \\
    Multiview + CL & frontal & 0.723 & 0.814 & 0.602 & 0.761 & 0.823 \\
    Multiview + CL + TL & frontal & \textbf{0.756} & \textbf{0.829} & \textbf{0.636} & \textbf{0.786} & \textbf{0.846} \\
    \midrule
    \midrule
    Single-view & lateral & 0.856 & 0.954 & 0.807 & \textbf{0.895} & 0.959 \\
    Single-view + CL~\cite{luo2021medical} & lateral & 0.840 & 0.946 & 0.809 & 0.872 & 0.948 \\ \midrule
    Multiview & lateral & 0.844 & 0.951 & 0.800 & 0.870 & 0.956 \\
    Multiview + TL & lateral & 0.848 & 0.954 & 0.804 & 0.876 & 0.961 \\
    Multiview +~\cite{jimenez2019medical} & lateral & 0.837 & 0.945 & 0.779 & 0.870 & 0.949 \\
    Multiview +~\cite{jimenez2019medical} + TL & lateral & \textbf{0.857} & \textbf{0.960} & \textbf{0.819} & 0.885 & \textbf{0.969} \\
    Multiview + CL & lateral & 0.838 & 0.956 & 0.807 & 0.867 & 0.956 \\
    Multiview + CL + TL & lateral & 0.840 & 0.955 & 0.794 & 0.874 & 0.960 \\ \bottomrule
    \end{tabular}
\end{table}

\section{Conclusion}
In this work, we propose a novel multiview deep learning method for elbow fracture subtype classification from frontal and lateral view X-ray images. We leverage transfer learning by first pretraining two single-view models. Meanwhile, medical knowledge was quantified and incorporated in the training process through curriculum learning. The results show that our multiview model outperforms the compared methods, and we achieved improved results over the previously published curriculum training strategies. As future work, we plan to further integrate other domain knowledge with respect to different views and explore curriculum learning in the output space.

\subsubsection{Acknowledgements.} This work was supported in part by National Institutes of Health grants (1R01CA193603 and 1R01CA218405), the Stimulation Pilot Research Program of the Pittsburgh Center for AI Innovation in Medical Imaging and the associated Pitt Momentum Funds of a Scaling grant from the University of Pittsburgh (2020), and an Amazon Machine Learning Research Award.

\bibliographystyle{splncs04}
\bibliography{mybibliography.bib}

\end{document}